\documentclass[aps,prl,onecolumn,10pt,noshowpacs,noshowkeys, notitlepage,twoside,a4paper,groupaddress]
{revtex4-1}
\usepackage{amssymb, amsmath}
\usepackage{color}
\usepackage{graphicx}

\begin{document}
\title{Subdiffusion via dynamical localization induced  by thermal equilibrium fluctuations}
\author{Jakub Spiechowicz}
\affiliation{Institute of Physics, and  Silesian Center for Education and Interdisciplinary Research, University of Silesia,  41-500 Chorz{\'o}w, Poland}

\author{Jerzy {\L}uczka}
\email[Correspondence to J.{\L}. ]{(e-mail: jerzy.luczka@us.edu.pl)}
\affiliation{Institute of Physics, and  Silesian Center for Education and Interdisciplinary Research, University of Silesia,  41-500 Chorz{\'o}w, Poland}

\begin{abstract}
We reveal the mechanism of subdiffusion which emerges in a straightforward, one dimensional classical \emph{nonequilibrium} dynamics of a Brownian ratchet driven by both a time-periodic force and Gaussian white noise.
In a tailored parameter set for which the deterministic counterpart  is in a \emph{non-chaotic} regime, subdiffusion is a long-living transient whose lifetime can be many, many orders of magnitude larger than characteristic time scales of the setup thus being amenable to experimental observations. As a reason for this  subdiffusive behaviour  in the coordinate space we identify thermal noise induced \emph{dynamical localization} in the velocity (momentum) space. This novel idea is distinct from existing knowledge and has never been reported for any classical or quantum systems. It suggests reconsideration of generally accepted opinion that subdiffusion is due to road distributions or strong correlations which reflect disorder, trapping, viscoelasticity of the medium  or geometrical constraints. 
\end{abstract}
%
%
\maketitle

Diffusion can be observed almost everywhere: in the material world (diffusion of particles, atoms, molecules, proteins, cytoplasmic macromolecules) \cite{mehrer,karger} and in the non-material world of human civilization at various levels of society organizations (diffusion of ideas, opinions, innovations, price values) \cite{rogers2003}. A physical archetype of diffusion is a Brownian motion resulting from interaction of a particle with its environment \cite{hanggi100years}. In literature one can find several quantifiers which characterize a diffusion process 
and spread of trajectories. An example is the mean-square displacement of the particle coordinate. In this paper, we will consider the  mean-square deviation (variance) of the particle position $x(t)$ around its mean value, namely, 
\begin{equation}
	\label{msd}
	\sigma_x^2(t) = \langle [ x(t) - \langle x (t) \rangle ]^2 \rangle, 
\end{equation}
where the averaging is over all thermal realizations as well as over initial conditions. The diffusion process can be classified through the scaling function \cite{zaburdaev2015}
\begin{equation}
	\label{scaling}
	\sigma_x^2(t) \sim  t^{\alpha}.  
\end{equation}
The normal diffusion corresponds to the scaling index $\alpha = 1$. Any deviation from this linear time dependence is classified as anomalous diffusion. For the superdiffusive case, 
$\sigma_x^2(t)$  increases over time faster while for the subdiffusion it grows slower than for normal diffusion. An example of the former is ballistic diffusion with the scaling index $\alpha = 2$. The hallmark of the latter is famous Sinai subdiffusion which follows the logarithmic law $\sigma_x^2(t) \sim \ln^4{t}$ \cite{sinai1982}. This ultraslow process can be observed for a Brownian particle moving in a static random Gaussian force field imitating quenched disorder in heterogeneous media. "Quenched" means that random traps, barriers or comb-like structures do not evolve with time. This is usually the model considered to describe the dynamical properties of materials containing impurities, defects, or intrinsic randomness like it is the case for amorphous systems \cite{buszo1990}. However, recent progress in single particle tracking techniques \cite{metzler2014} has allowed to probe transport processes occurring in more complex setups. For instance, the diffusive motion of macromolecules and organelles inside living cells is typically subdiffusive \cite{banks,regner}. This behaviour is commonly attributed to macromolecular crowding of their interior, summarizing their densely packed, heterogeneous and \emph{fluctuating} environment \cite{barkai2012, hofling2013}. In this new class of systems subdiffusion may be only a transient effect while  normal diffusion is observed in the asymptotic long time regime \cite{saxton2007}. Nevertheless this anomaly lasts sufficiently long for experimental detection \cite{platani2002, murase2004, bronstein2009, jeon2011}. Despite the outlined fundamental differences researchers try to exploit the well known mechanisms to capture the essence of anomalous diffusion (in the biological context, see e.g. Ref. \cite{berry}). They are either broad distributions or strong correlations in diffusive motion. Among others there are two major subdiffusive models. The first one is a system where the particle dynamics is found to be governed by a sequence of trapping-untrapping events, e.g. in energy wells. It is described in the framework of continuous time random walk \cite{zaburdaev2015} consisting of jump models where the particle undergoes a series of displacement given in terms of distribution of waiting times with power law tails. The second one is a system where the particle does not simply move in a fixed potential as before, but is part of an interacting setup exhibiting viscoelastic behaviour meaning that  dynamics of  different components of the system are correlated. This situation is described in terms of  fractional Brownian motion or a fractional Langevin equation \cite{eliazar2013, sokolov2012, meroz2015} by relaxing the white noise assumption in  simple Brownian motion and \emph{a priori} inclusion of a power law for time correlations of thermal noise. A more interesting situation occurs when this diffusion anomaly is induced by dynamics of the problem itself and is not related to genuine disorder or is not introduced \emph{ab initio} with broad distributions or strong correlations. 
We can mention models of anomalous diffusion (both subdiffusion and superdiffusion) in discrete deterministic chaotic systems 
 \cite{geisel1984,geisel85}. In contrast, here we use a bottom up approach: by analysing the Langevin equation for a Brownian ratchet driven by thermal white noise, we reveal a new mechanism how subdiffusion may emerge from deterministic dynamics superimposed with thermal noise. 
\begin{figure}[t]
	\centering
	\includegraphics[width=1.0\linewidth]{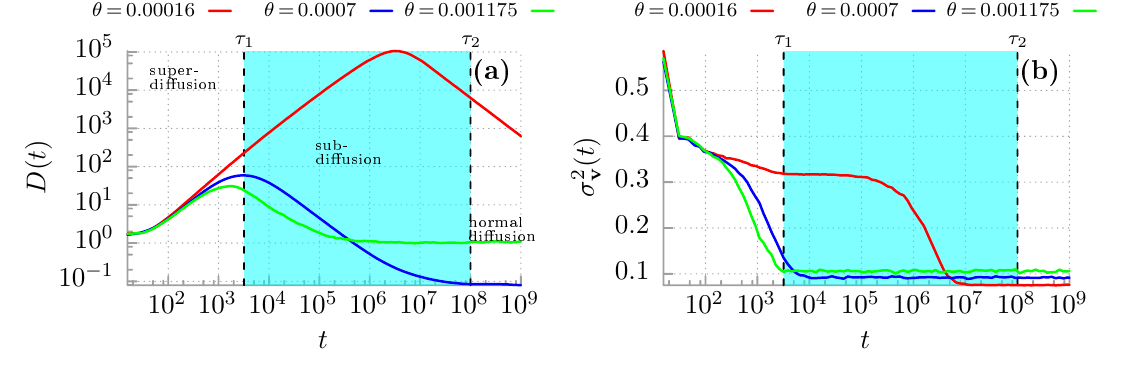}
	\caption{Diffusion anomalies of an inertial Brownian particle moving in a periodic potential and driven by a unbiased time-periodic force. (a) The time dependence of the diffusion coefficient $D(t)$. (b) The evolution of variance of the period averaged velocity $\sigma_\mathbf{v}^2(t)$ is presented for three values of thermal noise intensity $\theta$ proportional to temperature. The region corresponding to the subdiffusive behaviour for $\theta = 0.0007$ is indicated with the cyan colour. Parameters are $m = 6$, $a = 1.899$, $\omega = 0.403$. At zero temperature $\theta =0$, the system is non-chaotic.}
	\label{fig1}
\end{figure}

\section*{Results}
We consider an archetype of the Brownian motor \cite{hanggi2009, cubero2016} consisting of an inertial Brownian particle moving in a one-dimensional periodic potential of a ratchet type (i.e. with broken reflection symmetry) and driven by an external unbiased time-periodic force.  Its dynamics is determined  by the following dimensionless Langevin equation \cite{spiechowicz2016scirep}
\begin{equation}
	\label{model}
m\ddot{x} + \dot{x} = -U'(x) + a\cos{(\omega t)} + \sqrt{2\theta}\,\xi(t). 
\end{equation}
We refer the reader to the section Methods where we describe in detail the scaling procedure. The dot and the prime denote differentiation with respect to time $t$ and the Brownian particle coordinate $x \equiv x(t)$, respectively. The dimensionless friction coefficient is $1$ and the parameter $m$ is the dimensionless mass of the particle  moving in a spatially periodic potential $U(x) = U(x + L)$ of period $L$ and driven by the external unbiased time-periodic deterministic force $F(t)= a\cos{(\omega t)}$ of amplitude $a$ and angular frequency $\omega$. Thermal fluctuations due to coupling of the particle with the thermal bath of dimensionless temperature $\theta$ are modelled by $\delta$-correlated Gaussian white noise $\xi(t)$ of zero mean and unit intensity, i.e., 
\begin{equation}
	\label{noise}
	\langle \xi(t) \rangle = 0, \quad \langle \xi(t)\xi(s) \rangle = \delta(t-s).
\end{equation} 
The potential is assumed to be in the ratchet (asymmetric) form \cite{hanggi2009} 
\begin{equation}
	\label{potential}
	U(x) = -\sin{x} - \frac{1}{4}\sin{2x}
\end{equation}
of the period $L = 2\pi$. 

Due to the presence of the external driving \mbox{$F(t) = a \cos(\omega t)$} and the friction term $\dot x$ the particle velocity approaches for $t\to\infty$ a unique nonequilibrium stationary state which is characterized by a temporally periodic probability density. Then the mean velocity $\langle \dot{x}(t) \rangle$ takes the form of a Fourier series over all possible harmonics \cite{jung1993}
\begin{equation}
	\lim_{t \to \infty} \langle \dot{x}(t) \rangle = \langle \mathbf{v} \rangle + v_\omega(t) + v_{2\omega}(t) + ...,
\end{equation}
where $\langle \mathbf{v} \rangle$ is the directed (time independent) velocity while $v_{n\omega}(t)$ denote harmonic functions of vanishing average over the fundamental period $T = 2\pi/\omega$. Due to this particular decomposition it is useful to study the period averaged velocity $\mathbf{v}(t)$ defined as
\begin{equation}
	\label{velocity}
	\mathbf{v}(t) = \frac{1}{T} \int_t^{t + T} \dot{x}(s) \, ds
\end{equation}
which may be exploited to evaluate the directed velocity as 
$ \langle \mathbf{v} \rangle = \lim_{t \to \infty} \langle \mathbf{v}(t) \rangle$. A sufficient and necessary condition for the emergence of the directed transport $\langle \mathbf{v} \rangle \neq 0$  is breaking of the mirror symmetry of the potential $U(x)$ which is the case for the form described by Eq. (\ref{potential}) \cite{hanggi2009}. This operating principle can be seen as a key for understanding the intracellular transport \cite{bressloff2013}. Despite of simplicity of this system, it exhibits a number of notable and unusual features \cite{machura2007, spiechowicz2014pre, reimann2001, lindner2016, spiechowicz2016njp, spiechowicz2017chaos} including the anomalous diffusion \cite{spiechowicz2015pre,spiechowicz2016scirep}.
The origin of observed superdiffusion has already been satisfactorily explained in Ref. \cite{spiechowicz2016scirep}. Unfortunately, the mechanism standing behind subdiffusion still lacks a rewarding illumination. Within this work we aim to fill this major gap.
%

In panel (a) of Fig. \ref{fig1} we depict the 
time-dependent "diffusion coefficient" $D(t)$ defined as
\begin{equation}
    \label{D}
    D(t) =  \sigma_x^2(t)/2t.
\end{equation}
From Eq. (\ref{scaling}) it follows that superdiffusion occurs when $D(t)$ is an increasing function of time, the case of decreasing $D(t)$ corresponds to subdiffusion and for non-varying $D(t)$ normal diffusion takes place. The evolution of $D(t)$ can be divided into three time-domains, c.f. the blue curve for temperature $\theta=0.0007$ in panel (a): the early period of superdiffusion in the time-interval $(0, \tau_1)$, the intermediate regime of subdiffusion in the window $(\tau_1, \tau_2)$ shaded with the cyan colour and the asymptotic long time regime of normal diffusion for $t > \tau_2$. The crossover times $\tau_1$ and $\tau_2$ separating these domains can be \emph{controlled by temperature} and are monotonically decreasing function of thermal noise intensity \cite{spiechowicz2016scirep}. For example, when $\theta = 0.0007$ (the blue curve in Fig. 1), $\tau_1 \approx 3.2 \cdot 10^3$ and $\tau_2 \approx 10^8$. If temperature is lowered to $\theta = 0.00016$ (the red curve in Fig. 1) the superdiffusion lifetime is extended to $\tau_1 \approx 3.2 \cdot 10^6$. It is difficult to numerically determine $\tau_2$ due to limited stability of the utilized algorithm leading to uncontrolled propagation of roundoff and truncation errors. However, if we adopt a more reserved  extrapolation from the case  $\theta = 0.0007$, the time $\tau_2$  is at least of order $10^{11} \sim 10^{13}$. So,  these two times $\tau_1$ and $\tau_2$ are many, many orders longer than characteristic time scales of the system which in the dimensionless form are of the order of unity! Therefore from the experimental point of view for tailored parameter regimes these diffusion anomalies may be safely treated as nearly persistent effects.
\begin{figure}[t]
	\centering
	\includegraphics[width=1.0\linewidth]{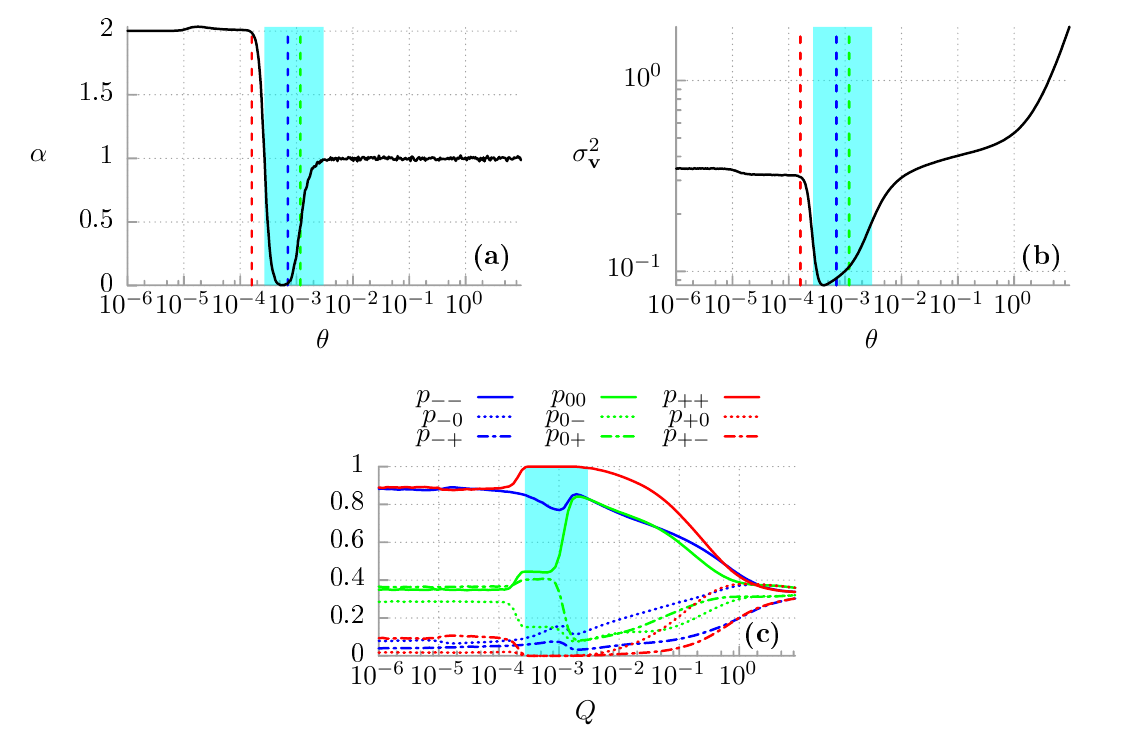}
	\caption{(a): The scaling index $\alpha$ vs temperature $\theta$ fitted to the asymptotic parts of  $\sigma_x^2(t)$. The temperature  interval corresponding to subdiffusion is indicated with the cyan colour. Superdiffusion is for lower temperatures ($\alpha >1$)  and normal diffusion - for higher temperatures ($\alpha = 1$). (b) The period averaged velocity variance $\sigma_\mathbf{v}^2(t)$ in the long time regime. The vertical lines indicate three values of $\theta =0.00016$ (red), $\theta =0.0007$ (blue) and $\theta =0.001175$ (green). (c) The conditional probabilities in the three-states model: the minus $v_- \approx -0.4$, the zero $v_0 \approx 0$ and the plus $v_+ \approx 0.4$ solution all presented as function of temperature $\theta$ of the system.  Other parameters are the same as in Fig. \ref{fig1}.}
	\label{fig2}
\end{figure}

To explain the mechanism standing behind the subdiffusion in such a periodic system let us now study the period averaged velocity $\mathbf{v}(t)$ of the Brownian particle, c.f. formula (\ref{velocity}) in the section Methods.
Its variance $\sigma_\mathbf{v}^2(t) = \langle \mathbf{v}^2(t) \rangle - \langle \mathbf{v}(t) \rangle^2$ characterises  fluctuations of the actual period averaged velocity around its mean value and is depicted in Fig. \ref{fig1} (b) for selected values of temperature $\theta$. On its basis, we can make two observations. The first is that the crossover time $\tau_1$ of superdiffusion is related to the relaxation of the period averaged velocity $\mathbf{v}(t)$ to its stationary state which for low to mid temperature may be ultraslow \cite{spiechowicz2016scirep}. The second is that when the  particle subdiffuses then the variance of the period averaged velocity $\sigma_\mathbf{v}^2(t)$ is much smaller than it is for superdiffusion. This fact suggests that during subdiffusion the coherence of motion is enhanced not only in the coordinate space but also in the velocity space. It is better visualized in Fig. \ref{fig2} (b) where we present the asymptotic value of the period averaged velocity variance $\sigma_\mathbf{v}$ computed for the final moment of time $t_f = 10^6$ in our numerical simulations as a function of temperature $\theta$. The interval corresponding to subdiffusion is shaded with the cyan colour, c.f. panel (a) of the same figure where we depict the scaling index $\alpha$ fitted to the asymptotic parts of the mean square deviation  of the particle coordinate $\sigma_x^2(t)$ evolved up to the same $t_f$. The variance is more than three times smaller in this temperature window than for values describing superdiffusion and normal diffusion. Moreover, the scaling index $\alpha$ can be controlled by temperature of the system in the non-monotonic way. There is a window of thermal noise intensity $\theta \in [7 \cdot 10^{-4}, 8.5 \cdot 10^{-4}]$ for which it is very small but still non-zero $\alpha \neq 0$ indicating nearly coherent transport or the ultraslow subdiffusion. For slightly lower and higher temperature the asymptotic evolution of $\sigma_x^2(t)$ is essentialy subdiffusive, e.g. $\alpha = 0.5$ for $\theta = 1.175 \cdot 10^{-3}$, c.f. Fig. \ref{fig2} (a). These findings suggest the prominent role which is played by thermal noise in the observed subdiffusive behaviour.

We further examine it by constructing a rough-and-ready three state stochastic process. In the deterministic limit $\theta = 0$ the system is \emph{non-chaotic} and possesses three coexisting attractors for the period averaged velocity: $v_- \approx -0.4$, $v_0 \approx 0$ and $v_+ \approx 0.4$ \cite{spiechowicz2016scirep}. Therefore we now analyse transitions between these states induced by thermal fluctuations. We introduce the following notation: $p_{++}$ stands for the (conditional) probability to remain staying in the plus state $v_+ \to v_+$, $p_{+-}$ denotes the probability for a jump between the opposite states $v_+ \to v_-$ and $p_{+0}$ is the probability of a transition between the plus and the zero state $v_+ \to v_0$. This convention is analogous for the remaining six  probabilities $\{p_{00}, p_{0+}, p_{0-}, p_{--}, p_{-0}, p_{-+}\}$. By introducing the threshold $\nu = 0.2$ the above probabilities can be estimated from simulations of the Langevin equation (\ref{model}) as the relative frequencies with which the period averaged velocity $\mathbf{v}(t)$ visits the three coarse grained regions $V_+ = \{|\mathbf{v}(t) - 0.4| < \nu \}$, \mbox{$V_0 = \{|\mathbf{v}(t)| < \nu \}$} and $V_- = \{|\mathbf{v}(t) + 0.4| < \nu \}$. A chosen value of the threshold $\nu = 0.2$ follows naturally from inspection of the probability distribution of the period averaged velocity where for low to moderate temperature regimes there are three peaks corresponding to the deterministic coexisting attractors which are approximately $\nu = 0.2$ apart (not depicted). Fig. \ref{fig2} (c) presents all transition probabilities as a function of temperature $\theta$. In the low temperature regime the probabilities for the particle to survive in the opposite running states $p_{++}$ and $p_{--}$ corresponding to the velocities $v_+ \approx 0.4$ and $v_- \approx -0.4$, respectively, are particularly large and almost equal. On the other hand, the probability for staying in the locked state $p_{00}$ is twice lower. Consequently, the spread of trajectories is large and ballistic motion is observed. The temperature interval for which subdiffusion is developed is again marked by the cyan colour. There thermal noise induces dynamical localization  of the period averaged velocity (momentum), i.e. it resides in the positive running state $v_+$ with the probability $p_{++}$ very close to unity. It means that once the ensemble of particles enter this state in the velocity space it moves almost coherently, i.e. with marginal fluctuation allowed in the chosen threshold $|\mathbf{v}(t) - 0.4| < \nu$. These fluctuations are responsible for the ultraslow subdiffusion where the observed scaling index is very small but still nonzero $\alpha \neq 0$. We also note that in this interval the probability for surviving in the locked state $p_{00}$ corresponding to the particle locked in one of the potential wells is enlarged. 
Each stay of the particle in the locked state hampers the increase of the spread of trajectories leading to slowing down of diffusion. Similar mechanism contributes to subdifussion observed in systems with disorder, e.g. random barriers, where the particle dynamics is governed by trapping-untrapping events in energy wells. However, certainly the dominant factor which is responsible for the detected subdiffusion is dynamical localization in the positive running state as it is depicted in Fig. \ref{fig2} (c). Further increase of temperature immediately delocalizes the particle. However, since then the probability for being in the locked state $p_{00}$ is at the level of $p_{--}$ and noticeable smaller than $p_{++}$ we observe the occurrence of normal diffusion. In the high temperature limit all transition probabilities are almost the same.
\begin{figure}[t]
	\centering
	\includegraphics[width=1.0\linewidth]{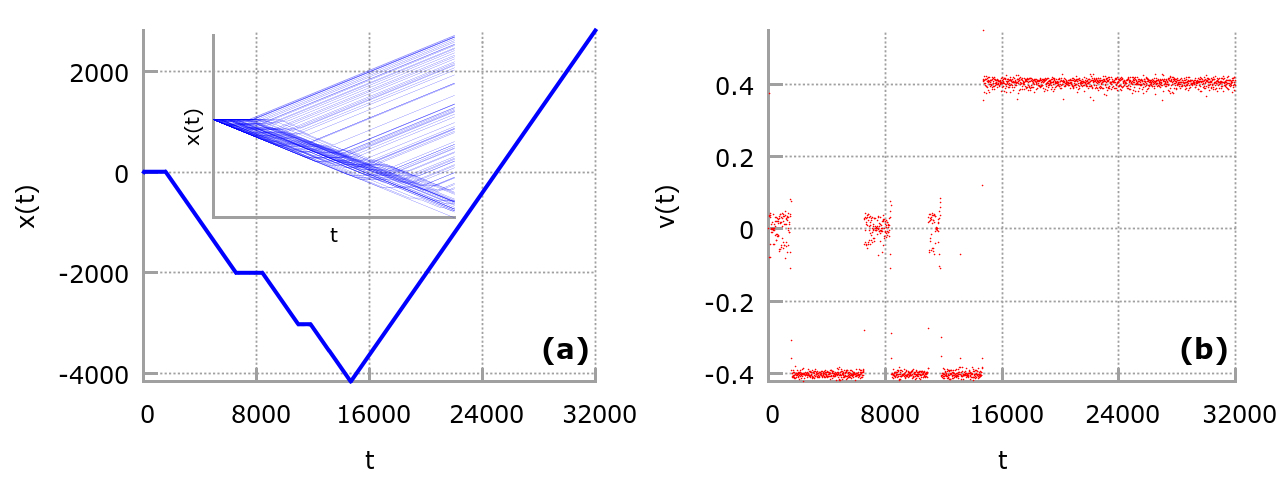}
	\caption{A representative trajectory of the Brownian particle coordinate $x(t)$ and the period averaged velocity $\mathbf{v}(t)$. Each red dot in panel (b) depicts the period averaged velocity $\mathbf{v}(t)$ for a given $t$. The dynamical localization of the latter in the state pointing to the positive direction $v_+ \approx 0.4$ is illustrated. In the inset of panel (a) we show the ensemble of 100 trajectories of the Brownian particle. Parameters are the same as in Fig. \ref{fig1} with $\theta = 0.0004$.}
	\label{fig3}
\end{figure}

To better visualize the explained mechanism we now focus on the single representative Brownian particle trajectory for temperature $\theta = 0.0004$ corresponding to subdiffusive behaviour. The result is shown in Fig. \ref{fig3}. In panel (a) and (b) we present the evolution of the coordinate $x(t)$ and the period averaged velocity $\mathbf{v}(t)$, respectively. Each red dot in panel (b) depicts the latter quantity for a given $t$. The time intervals corresponding to motion in each of the observed states $v_-$, $v_0$, $v_+$ can be easily identified. Since in this regime $p_{++} \approx 1$ once the particle enters the plus state it stays in its vicinity $\mathbf{v}(t) = v_+ \approx 0.4$ (c.f. panel (b)) for very long time. After the dynamical localization all particles travel with almost identical velocity and the spread of their trajectories changes very little over time implying the subdiffusive behaviour. We illustrate this mechanism in the inset of \mbox{Fig. \ref{fig3} (a)} where we depict an ensemble of 100 trajectories of Brownian particle dynamics for the same temperature. However, we stress that the magnitude of the estimated probability for surviving in the plus state $p_{++}$ naturally depends on the chosen threshold $\nu$ of the velocity space coarse graining procedure. We restricted ourselves to the simplest stochastic model capturing the essence of the mechanism standing behind the observed subdiffusion.  Summing up this part, for the fixed temperature $\theta$ and small times $t < \tau_1$ the dynamics of each trajectory is mostly deterministic giving rise to the (ballistic) superdiffusive behaviour provided that it is averaged over unbiased initial conditions for the  particle position and velocity. The escape rates from the minus and zero attractors are related to the crossover time $\tau_1$ of the superdiffusive stage of motion. For intermediate time regime $\tau_1 < t < \tau_2$ thermal fluctuations induce transitions from the minus and the zero states onto the plus velocity. The observed dynamical localization in the plus state means that the typical escape rate from this solution - which is related to the crossover time $\tau_2$ of subdiffusion - is much larger than the corresponding quantity for two other  attractors thus eventually luring there almost all trajectories and hindering the diffusion rate. We numerically verify that for temperature $\theta \approx 0.0007$ for which the power exponent $\alpha \approx 0$ in the interval $\tau_1 < t < \tau_2$ the probability for the particle to be in the plus state is very, very close to one $p_+ \approx 1$ meaning that almost all particles are localized in this solution. 
Finally, for sufficiently long times $t > \tau_2$ random dynamics induced by thermal fluctuations activates escape from all attractors and jumps between various trajectories leading to  normal diffusion. 


\section*{Discussion}
In conclusion, we explained the mechanism of long lasting subdiffusion in a nonequilibrium noisy system, which in the deterministic limit is non-chaotic. 
As the reason for this anomalous behaviour we identified  classical dynamical localization \cite{guarneri2014} of the particle velocity (momentum) induced by thermal equilibrium fluctuations, i.e. there is a temperature window for which the particle localizes in the positive running state $v_+$ with the probability very close to unity. This temperature interval matches exactly the region where the subdiffusive regime is observed in the finite data acquisition time. 

Dynamical localization is a well known phenomenon in quantum systems and was introduced in Ref. \cite{chirikov}. It manifests itself in quantum suppression of the chaotic classical diffusion in momentum space due to interference effects \cite{bitter}. A different but closely related phenomenon to dynamical localization in deterministic systems is Anderson spatial localization in disordered systems \cite{anderson,fiszman}. 
With this work we established a relation between subdiffusion in the coordinate space and the dynamical (quasi)localization in the momentum space. To the best of our knowledge, a similar idea has never been reported for any other classical or quantum systems.  
The findings are distinct from existing knowledge and suggest reconsideration of generally accepted views that subdiffusion is due to  broad distributions or strong correlations which reflect disorder, trapping, viscoelasticity of the medium or geometrical constraints. 

Since we demonstrated the new mechanism of subdiffusion using the model of a Brownian ratchet, the results provide essentially new insight into processes governing transport in complex systems especially of biological origin. Notably, they  may explain certain aspects of strange kinetics occurring in living cells, with a particular emphasis on subdiffusive motion of molecular motors which are responsible for the intracellular transport. 
While we illuminate this phenomenon in the case of the driven Brownian particle moving in the asymmetric (ratchet) potential, in principle there are no restrictions for this \emph{universal} mechanism to be likely observed also for other nonequilibrium setups: driven or non-driven but tilted symmetric periodic systems \cite{reimann2001,lindner2016} and including those which operate in strong disspation regime or in overdamped limit. The latter is especially true due to the fact that in the deterministic limit our setup is in a non-chaotic regime. Therefore complex chaotic dynamics which is characteristic in one dimension for driven inertial systems is not needed for this mechanism to occur. Moreover, due to simplicity and universality of the system with its physical clarity as well as appealing strength of Brownian motion with its intrinsic Gaussian noise propagator our findings can open a wide area of studies and may be corroborated experimentally with a wealth of physical systems. One of the most promising setups for this purpose are optical lattices \cite{renzo,lutz2013,lutz2017,wickenbrock2012} and asymmetric SQUID devices \cite{spiechowicz2014prb,spiechowicz2015njp}.
\section*{Methods}
In this work we analyse the generic model of a ratchet system which consists of (i) a classical inertial particle of mass $M$, (ii) moving in a deterministic asymmetric ratchet potential $U(x)$, (iii) driven by an unbiased time-periodic force $A\cos{(\Omega t)}$ of amplitude $A$ and angular frequency $\Omega$, and (iv) subjected to thermal noise of temperature $T$. The corresponding Langevin equation reads \cite{spiechowicz2016scirep}
\begin{equation} \label{LL}
	M\ddot{x} + \Gamma\dot{x} = -U'(x) + A\cos{(\Omega t)} + \sqrt{2\Gamma k_B T}\,\xi(t),
\end{equation}
The parameter $\Gamma$ stands for the friction coefficient and $k_B$ is the Boltzmann constant. Thermal equilibrium fluctuations are modeled by $\delta$-correlated, Gaussian white noise $\xi(t)$ of zero mean and unit intensity, i.e.,
\begin{equation}
	\langle \xi(t) \rangle = 0, \quad \langle \xi(t)\xi(s) \rangle = \delta(t-s).
\end{equation}
The spatially periodic potential $U(x)$ is assumed to be in a double-sine form of period $2\pi L$ and a barrier height $\Delta U$, namely
\begin{equation} \label{pot} 
	U(x) = -\Delta U\left[ \sin{\left(\frac{x}{L}\right)} + \frac{1}{4}\sin{\left( 2 \frac{x}{L} + \varphi - \frac{\pi}{2}\right)}\right].
\end{equation}

As only relations between scales of length, time and energy are relevant but not their absolute values we next formulate the above equations of motion in its dimensionless form. To do so, we first introduce the characteristic dimensionless scales for the system under consideration
\begin{equation} \label{scales} 
	\hat{x} = \frac{x}{L}, \quad \hat{t} = \frac{t}{\tau_0}, \quad \tau_0 = \frac{\Gamma L^2}{\Delta U},
\end{equation}
so that the dimensionless form of the Langevin dynamics (\ref{model}) reads
\begin{equation}
	\label{dimlessmodel}
	m\ddot{\hat{x}} + \dot{\hat{x}} = -\hat{U}'(\hat{x}) + a\cos{(\omega \hat{t})} + \sqrt{2Q} \hat{\xi}(\hat{t})\;.
\end{equation}
Here, the dimensionless potential $\hat{U}(\hat{x}) = U(x)/\Delta U = U(L\hat{x})/\Delta U = \hat{U}(\hat{x} + 2\pi)$ possesses the period $2\pi$ and the unit barrier height is $\Delta U = 1$. Other parameters are: $m = M/(\Gamma\tau_0)$, $a = (L/\Delta U)A$, $\omega = \tau_0\Omega$. The rescaled thermal noise reads \mbox{$\hat{\xi}(\hat{t}) = (L/\Delta U)\xi(t) = (L/\Delta U)\xi(\tau_0\hat{t})$} and assumes  the same statistical properties, namely $\langle \hat{\xi}(\hat{t}) \rangle = 0$ and \mbox{$\langle \hat{\xi}(\hat{t})\hat{\xi}(\hat{s}) \rangle = \delta(\hat{t} - \hat{s})$}. The dimensionless noise intensity $Q = k_BT/\Delta U$ is the ratio of thermal and the activation energy the particle needs to overcome the nonrescaled potential barrier.

The system described by Eq. (\ref{LL}) possesses four characteristic time scales
\begin{equation}
	\label{tscales} 
	\tau_0 = \frac{\Gamma L^2}{\Delta U}, \qquad \tau_1 = \frac{M}{\Gamma}, \qquad \tau_2^2 = \frac{ML^2}{\Delta U}, \qquad \tau_3 = \frac{2\pi}{\Omega}.  
	\end{equation}

The quantity $\tau_0$ is used as a unit of time in our scaling procedure, see Eq. (\ref{scales}). It is a characteristic time for an overdamped particle to move from the maximum of the potential $U(x)$ to its minimum. The scale $\tau_1$ is a relaxation time of the velocity of the free Brownian particle (when $U(x)=A=0$). 
Note that the dimensionless mas $m=\tau_1/\tau_0$ is a ratio of the two characteristic times.   The time $\tau_2$ is a characteristic scale for the conservative system (when $\Gamma=A=0$). The last scale $\tau_3$ is a period of the external time-periodic force. Thermal fluctuations are modelled here approximately as white noise so its correlation time is zero and there is no characteristic time scale associated with it. However, in real systems it is non-zero but much, much smaller than the other characteristic time scales.   
	
The Langevin equation (\ref{LL}) has been originally derived for the asymmetric superconducting quantum interference device (SQUID) which is composed of a loop with three capacitively and resistively shunted Josephson junctions, see Eq. (14) in Ref.  \cite{spiechowicz2014prb}. The particle coordinate $x$ and velocity $v$ corresponds to the Josephson phase and the voltage drop across the device, respectively. The particle mass stands for the capacitance of the SQUID, the friction coefficient translates to the reciprocal of the SQUID resistance. The time-periodic force corresponds to the external current. The asymmetry parameter $\varphi$ of the potential (\ref{pot}) can be controlled by the external magnetic flux which pierces the device.

The Langevin equation (\ref{dimlessmodel}) is solved by employing a weak version of the stochastic second-order predictor corrector algorithm and using a CUDA environment implemented on a modern desktop GPU. This procedure allows for a speedup of a factor of the order $10^3$ times as compared to a common present-day CPU method. Details on this implementation can be found in \cite{spiechowicz2015cpc}.  Since Eq. (\ref{dimlessmodel}) is a second-order differential equation we need to specify two initial conditions: one for the position $x(0)$ and the other one for the velocity $\dot{x}(0)$ of the Brownian particle. For some regimes of system parameters the studied dynamics may depend on a specific choice of these initial conditions and therefore to avoid this unwanted behaviour we chose $x(0)$ and $\dot{x}(0)$ to be equally distributed over the intervals $[0, 2\pi]$ and $[-2,2]$. However, we note that the effects discussed in the paper are robust with respect to variation of these initial conditions. In particular, in Ref. \cite{spiechowicz2016scirep} we have considered  three various forms of the probability distribution for the initial velocity of the particle: 
the  Dirac delta $P(v) = \delta(v)$, uniformly distributed $P(v) = U(-2,2)$ on the velocity interval $[-2, 2]$ and normally distributed (Gaussian)  $P(v) = N(0,1)$ of zero mean  and  standard deviation equals to one. All considered forms lead to the consistent results proving that the observed behaviour is robust with respect to them.
\section*{Acknowledgement}
The work was supported by the Grant NCN 2015/19/B/ST2/02856.
\section*{Author Contributions}
J.S. performed the numerical simulations. Both authors contributed extensively to the planning, interpretation, discussion and writing up of this work.
\section*{Competing financial interests}
The authors declare no competing financial interests.

\end{document}